\documentclass[aps,prl,twocolumn,showpacs,superscriptaddress]{revtex4-1}
\usepackage{graphicx}
\usepackage{amsmath}
\usepackage{amssymb}
\usepackage{psfrag}
\usepackage{natbib}
\usepackage{color,xcolor}
\begin{document}
\title{Solitary synchronization waves in distributed oscillators populations}
\author{L. A. Smirnov}
\affiliation{Department of Control Theory, Nizhny Novgorod State University,
Gagarin Av. 23, 606950, Nizhny Novgorod, Russia}
\affiliation{Institute of Applied Physics, Ul'yanov str.{\,}46, 603950, 
Nizhny Novgorod, Russia}
\author{G. V. Osipov}
\affiliation{Department of Control Theory, Nizhny Novgorod State University,
Gagarin Av. 23, 606950, Nizhny Novgorod, Russia}
\author{A. Pikovsky}
\affiliation{Institute for Physics and Astronomy, University of Potsdam, 
Karl-Liebknecht-Str. 24/25, D-14476 Potsdam-Golm, Germany}
\affiliation{Department of Control Theory, Nizhny Novgorod State University,
Gagarin Av. 23, 606950, Nizhny Novgorod, Russia}
\begin{abstract}
We demonstrate existence of solitary waves of synchrony in one-dimensional arrays of
identical oscillators with Laplacian coupling.
Coarse-grained description of the array leads to nonlinear equations for the complex
order parameter, in the simplest case to lattice equations similar to those of the
discrete nonlinear Schr\"odinger lattice.
Close to full synchrony, we find solitary waves for the order parameter perturbatively,
starting from the known phase compactons and kovatons; these solution are extended
numerically to the full domain of possible synchrony levels.
For non-identical oscillators, existence of dissipative solitons is demonstrated.
\end{abstract}
\maketitle
The dynamics of oscillator populations attracts a lot of interest across different
fields of science and engineering.
The paradigmatic and universal object of study is the Winfree-Kuramoto model of
globally coupled phase oscillators.
It demonstrates a transition to synchrony, characterized in the terms of the Kuramoto
order parameter~\cite{Kuramoto-75, *Acebron-etal-05}.
This effect is relevant for many systems (lasers, biocircuits, electronic and
electro-chemical oscillators~\cite{Nixon_etal-13, *Kiss-Zhai-Hudson-02a, *Temirbayev_et_al-12, *Prindle_etal-12})
which can be well-described within the mean-field coupling models.
In the case the oscillators are organised as an ordered medium or a lattice with a
distant-dependent coupling, spatio-temporal patterns can be observed.
Most popular here are standing chimera states~\cite{Panaggio-Abrams-15, *Omelchenko2018},
first reported for a one-dimensional~(1D) medium by  Kuramoto and Battogtokh~(KB)~\cite{Kuramoto-Battogtokh-02},
where regions of synchrony and asynchrony coexist.
The smallest system to observe chimera experimentally is that of 
two subpopulations~\cite{Tinsley_etal-12, *Martens_etal-13}.
Further experiments have been performed with media
of nonlocally coupled chemical oscillators~\cite{Nkomo_etal-13, *Totz_etal-18} 
(up to 1600 units).

While 1D chimera patterns typically are stationary solutions, size of which is a characteristic
system size, a possibility of localised traveling solitary waves of the complex order parameter
in oscillatory media remains an open problem. 
In this Letter we report on solitary synchronization waves in a 1D
oscillatory medium with \textit{Laplacian} coupling.
In fact, our model is very similar to the KB setup with an interaction between the
oscillators defined through a convolution integral.
The only difference is in the type of the interaction kernel: in the KB model it is
an exponential function with non-vanishing integral, thus ensuring a mean-field-type
coupling within a neighborhood.
We use a Laplacian kernel, integral over which vanishes.
Such a coupling, as we show below, provides a plethora of solitary wave solutions. 
The main tool of our analysis is based on the Ott-Antonsen~(OA)
ansatz~\cite{Ott-Antonsen-08}, allowing to write closed equations for the complex
order parameter field
(see~\cite{Bordyugov-Pikovsky-Rosenblum-10, *Laing-11, *Smirnov-Osipov-Pikovsky-17} for
applications of this ansatz to the KB-type chimeras).
In the simplest setup, we will obtain a lattice equation for the order parameter
(this system can be interpreted as a lattice generalization of systems of two and three
coupled
populations of identical oscillators, studied
in~\cite{Abrams-Mirollo-Strogatz-Wiley-08, *Pikovsky-Rosenblum-08,*Martens-Panaggio-Abrams-16,%
*Martens_etal-16,*Kotwal-Jiang-Abrams-17} and \cite{Martens-10a,*Martens-10b}, respectively)
resembling the nonlinear Schr\"odinger~(NLS) lattice~\cite{Eilbeck-Johansson-03}. 
We will find solitary waves via a perturbation method starting with compacton solutions
for the fully synchronous case, and will describe the full domain of existence of
localized waves for different levels of synchrony.
Furthermore, we will show that for 1D arrays with diversity of natural 
frequencies and with additional attractive coupling compensating this diversity, 
waves of synchrony exist as dissipative solitons.

We start with a formulation of our basic model in a discrete form, assuming $x\!=\!\ell{q}$,
where $\ell$ is the spacing between oscillators and $q$ is their integer index.
The model we employ is close to the KB model of a 1D medium of
nonlocally coupled oscillators~\cite{Kuramoto-Battogtokh-02}.
The oscillators are described by their phases $\varphi_{q}\!\left(t\right)$ that are
governed by
\begin{equation}\label{eq:phdiscr}
\!\frac{d\varphi_{q}}{dt}\!=\!
\mathrm{Im}\!\left(H_{q}\!\left(t\right)e^{-i\varphi_{q}}\right)\!, \hspace{2.5mm}
H_{q}\!\left(t\right)\!=\!e^{-i\alpha}\ell
\sum\limits_{\tilde{q}}\!{G_{q\tilde{q}}e^{i\varphi_{\tilde{q}}(t)}}.\!\!\!
\end{equation}
Here we assume all oscillators to be identical, so that their natural frequency
can be set to zero by virtue of a transformation to the rotating reference frame.
The coupling is defined as a sum over the neighbours, with an additional phase shift
$\alpha$ and a kernel $G_{q\tilde{q}}=G\!\left(\ell(q-\tilde{q})\right)$, which is
assumed to be either a localized or a rapidly decaying function. 
In the limit $\ell\!\to\!{0}$, the model~\eqref{eq:phdiscr} is formulated as
a continuous oscillatory medium with an integral coupling:
\begin{equation}\label{eq:phint}
\frac{\partial\varphi}{\partial{t}}\!=\!
\mathrm{Im}\!\left(He^{-i\varphi}\right), \hspace{2.5mm}
H(x,t)\!=\!e^{-i\alpha}\!\!\int\!\!
G\!\left(x-\tilde{x}\right)e^{i\varphi(\tilde{x},t)}d\tilde{x}\,.
\end{equation}
While Eq.~\eqref{eq:phint} is suitable for the theoretical analysis, 
in numerics we simulate its discrete analog~\eqref{eq:phdiscr}. 

In Ref.~\cite{Kuramoto-Battogtokh-02}, as well as in many other studies of chimera
patterns~\cite{Panaggio-Abrams-15, *Omelchenko2018}, the interaction was assumed
to be of a mean-field type, characterized by a nonzero mean strength
$\int{G(x)dx}\!\neq\!{0}$.
Here, in contradistinction, we choose a \textit{Laplacian} symmetric coupling with a
vanishing mean value, i.e. $\int{G(x)dx}\!=\!0$.
The prototypic example of such a kernel is $G\!\left(x\right)\sim\left(x^2-1\right)
e^{-x^{2}/2}$. 

The essential advantage of the continuous limit is that we can perform a
local averaging to define the coarse-grained complex order parameter $Z\!\left(x,t\right)=\langle e^{i\varphi(x,t)}\rangle$,
where the averaging is performed over a small neighborhood of site $x$.
For $Z\!\left(x,t\right)$ one can furthermore write the OA 
equation~\cite{Panaggio-Abrams-15, *Omelchenko2018, Ott-Antonsen-08}
\begin{equation}\label{eq:opint}
\frac{\partial{Z}}{\partial{t}}\!=\!\frac{1}{2}\!\left(H\!-\!H^{\ast}Z^{2}\right)\!,\hspace{2.5mm} 
H\!\left(x,t\right)\!=\!e^{-i\alpha}\!\int\!{G\!\left(x\!-\!\tilde{x}\right)\! Z\!\left(\tilde{x},t\right)\!d\tilde{x}}.
\end{equation}
The coarse-grained order parameter describes the level of local synchrony of the units:
$|Z|=0$ if the phases are uniformly distributed (full asynchrony), and $|Z|=1$ in the
case of full synchrony where the phases coincide.
Partial synchrony corresponds to $0<|Z|<1$.

We now make one more simplification and replace the integral continuous model with a lattice one.
The model which captures all the essential properties of Eq.~\eqref{eq:opint} is a
lattice with nearest-neighbour Laplacian coupling
\begin{equation} \label{eq:bm}
\frac{dZ_{n}}{dt}\!=\!\frac{1}{2}\!\left(H_{n}\!-\!H_{n}^{\ast}Z_{n}^{2}\right),\hspace{2.5mm}
H_{n}\!=\!e^{-i\alpha}\left(Z_{n-1}\!+\!Z_{n+1}\!-\!2Z_n\right).
\end{equation}
Here the spatial index $n$ is not related to the original index $q$ of the phases,
rather $n$ describes domains that contribute to the
coupling field $H$ as ``coarse-grained macroscopic lattice sites''.  
The model~\eqref{eq:bm} is most suitable for the numerical and analytical
analysis of solitary waves, however, we will also present numerical evidence of found
waves for the model~\eqref{eq:phdiscr}.
Noteworthy, the lattice model~\eqref{eq:bm} can be viewed as an exact OA description for a nearest-neighbour interacting network of phase oscillators, where at each site $n$
there is a large population of $M$ units (phases $\varphi_{nm}$)
which are fully characterized, according to the OA
theory, by the local complex order parameter $Z_n=M^{-1}\sum_{m=1}^M e^{i\varphi_{nm}}$.
Such a model in the case of two coupled populations is known as the Abrams et al. model,
possessing chimera states (see~\cite{Abrams-Mirollo-Strogatz-Wiley-08, *Pikovsky-Rosenblum-08,%
*Martens-Panaggio-Abrams-16,%
*Martens_etal-16,*Kotwal-Jiang-Abrams-17} and~\cite{Martens-10a,*Martens-10b}
for the theoretical analysis of two and tree populations,
and \cite{Tinsley_etal-12, Martens_etal-13} 
for experimental realizations).

The dynamics of the lattice~\eqref{eq:bm} crucially depends on shift $\alpha$, which 
determines whether the coupling is attractive, repulsive, or neutral.
We start with the neutral case $\alpha\!=\!-\pi/2$, where the dynamics of the lattice is conservative.
Spatially homogeneous solutions have the form $Z_{n}\!=\!\varrho e^{i\psi}$ with any
$0\leq\varrho\leq{1}$, i.e. any level of homogeneous synchrony is possible.
Linear waves $\propto{e^{iwt-ikn}}$ on top of such a homogeneous background
have dispersion
\begin{equation}\label{eq:dispersion}
w=\sqrt{1-\varrho^{4}}\left(1-\cos{k}\right).
\end{equation}

\looseness=-1
We now look for nonlinear solitary waves on top of a 
homogeneous background in lattice~\eqref{eq:bm} for $\alpha\!=\!-\pi/2$.
The only parameter is the homogeneous level of synchrony $\varrho$.
It is instructive to start with the degenerate case of full synchrony $\varrho\!=\!1$.
As it follows from~\eqref{eq:dispersion}, in this case there are no linear waves. 
In fact, because $|Z_{n}|\!=\!1$, the only nontrivial variable is the phase of the order
parameter, and the equation for this phase is the same as for a lattice of neutrally
coupled phase oscillators, studied in~\cite{Rosenau-Pikovsky-05, *Pikovsky-Rosenau-06,*Ahnert-Pikovsky-08}. 
With $Z_{n}\!=\!e^{i\Theta_{n}}$ and $V_{n}\!=\!\Theta_{n}\!-\!\Theta_{n-1}$, the dynamical equations
can be reduced to a simple lattice system
\begin{equation}\label{eq:vnCompactons}
\frac{dV_{n}}{dt}=\cos{V_{n+1}}-\cos{V_{n-1}}\;.
\end{equation}
Solitary waves in this fully synchronous lattice, compactons and kovatons, have been
thoroughly analysed in~\cite{Rosenau-Pikovsky-05, *Pikovsky-Rosenau-06,Ahnert-Pikovsky-08}, 
here we briefly outline their main features.
Traveling waves $V_n(t)\!=\!V(\tau)$, 
where $\tau\!=\!t\!-\!n/\lambda$,
can be either compact one-hump pulses with velocities $0\!<\!\lambda\!<\!{\lambda_{c}}\!=\!4/\pi$,
or extended domains between two compact kinks, connecting states $V\!=\!0$ and $V\!=\!\pi$, (kovatons) with velocity
$\lambda\!=\!\lambda_{c}$.

Next, in the framework of Eqs.~\eqref{eq:bm} we look for solitary waves moving with
constant velocities on top of a partially synchronous homogeneous background with
$\varrho\!<\!1$.
Substituting in~\eqref{eq:bm} $Z_{n}\!=\!\rho_{n}e^{i\theta_{n}}$, we obtain a system
\begin{subequations}
\label{eq:rhotheta}
\begin{align}
\!\!\!\frac{d\rho_{n}}{dt}\!&=\!\frac{\left(1-\rho_{n}^{2}\right)}{2}
\Bigl(\rho_{n-1}\sin{v_{n}}\!-\!\rho_{n+1}\sin{v_{n+1}}\Bigr),\label{eq:rho_n}\\
\!\!\!\frac{d\theta_{n}}{dt}\!&=\!\frac{\left(1+\rho_{n}^{2}\right)}{2\rho_{n}}
\Bigl(\rho_{n-1}\cos{v_{n}}\!+\!\rho_{n+1}\cos{v_{n+1}}\!-\!2\rho_{n}\Bigr),\label{eq:theta_n}
\end{align}
\end{subequations}
where  $v_{n}\!=\!\theta_{n}\!-\!\theta_{n-1}$.
We employ a traveling wave ansatz $\rho_n(t)\!=\!\rho(\tau)$,
$\theta_n(t)\!=\!\theta(\tau)$, and assume that $\rho\left(\tau\right)$ and
$\theta(\tau)$ satisfy conditions $\rho(-\infty)\!=\!\rho(+\infty)\!=\!\varrho$, 
$\theta(-\infty)\!=\!\theta^{-}$, $\theta(+\infty)\!=\!\theta^{+}$ ($\theta^{-}$ and $\theta^{+}$ are two constants).

\begin{figure}[!t]
\includegraphics[width=\columnwidth]{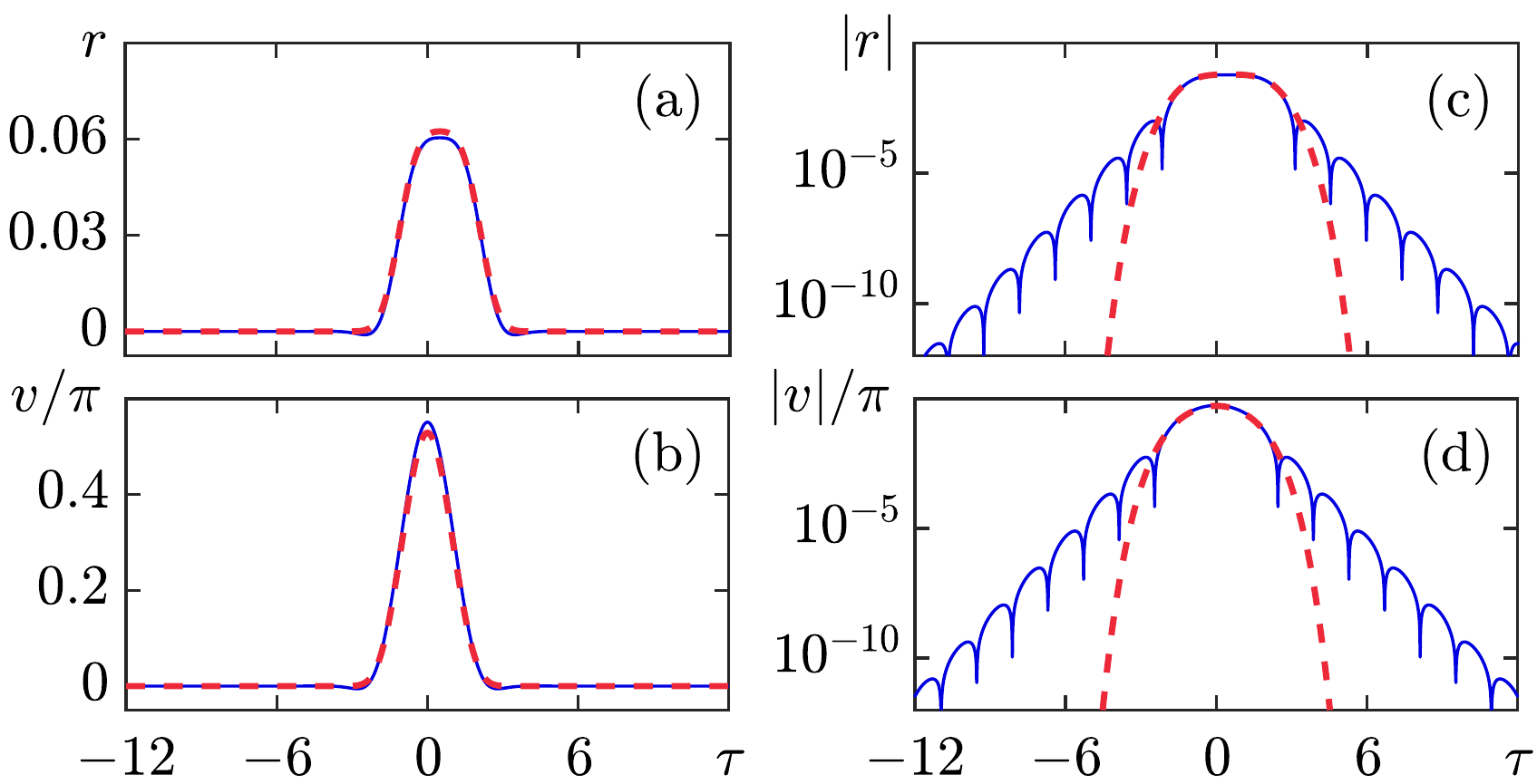}
\caption{Numerically obtained soliton for $\lambda\!=\!1.01862$ and $\varrho\!=\!0.9$ 
(solid blue line) compared with the approximation~\eqref{eq:apsol} (red dashed line). 
In the logarithmic scale (b,d) one can clearly see exponentially 
decaying tails with oscillations. }
\label{fig:solap}
\end{figure}

First, we develop a perturbation approach allowing to find solutions analytically for
the case close to synchrony $\varrho\!\lesssim\!{1}$.
Introducing a small parameter $\epsilon=(1-\varrho)\!\ll{1}$, we write
$\rho(\tau)\!=\!\varrho\!+\!{\epsilon}r_{1}(\tau)\!+\!o(\epsilon^{2})$,
$\theta(\tau)\!=\!\Theta(\tau)\!+\!\epsilon\vartheta_{1}(\tau)\!+\!o(\epsilon^{2})$.\\
Here $\Theta(\tau)\!=\!\Theta_n(t)$ is a compacton solution of~\eqref{eq:vnCompactons}.
Mostly important is the evolution of the correction ${\epsilon}r_1(\tau)$
to the constant value $\varrho$, so we consider only it below.
Substituting the expansion in $\epsilon$ to Eq.~\eqref{eq:rho_n}, we obtain an
expression, which allows us to represent $r_{1}\!\left(\tau\right)$ as an integral over
the compacton waveform
\begin{equation}\label{eq:apsol}
r_1(\tau)=1-\,\exp\!\left[\int_{-\infty}^{\tau}
\!\Bigl(\sin{V(\tilde{\tau}\!-\!\lambda^{-1})}\!-\!\sin{V(\tilde{\tau})}\Bigr)d\tilde{\tau}\right]\!.\!
\end{equation}
In this approximation the profile $r_1(\tau)$ is as compact as
the compacton solution of~\eqref{eq:vnCompactons}, {\,}i.e. it has superexponentially 
decreasing tails. 
The exact solitary wave solution of system~\eqref{eq:rhotheta} has exponentially
decaying tails like usual solitons -- because for $\varrho<1$ this system possess also
linear wave solutions.
We compare the approximate solution~\eqref{eq:apsol} with the numerical  solitary wave
in Fig.~\ref{fig:solap}.
  
\begin{figure}[!t]
\centering
\includegraphics[width=\columnwidth]{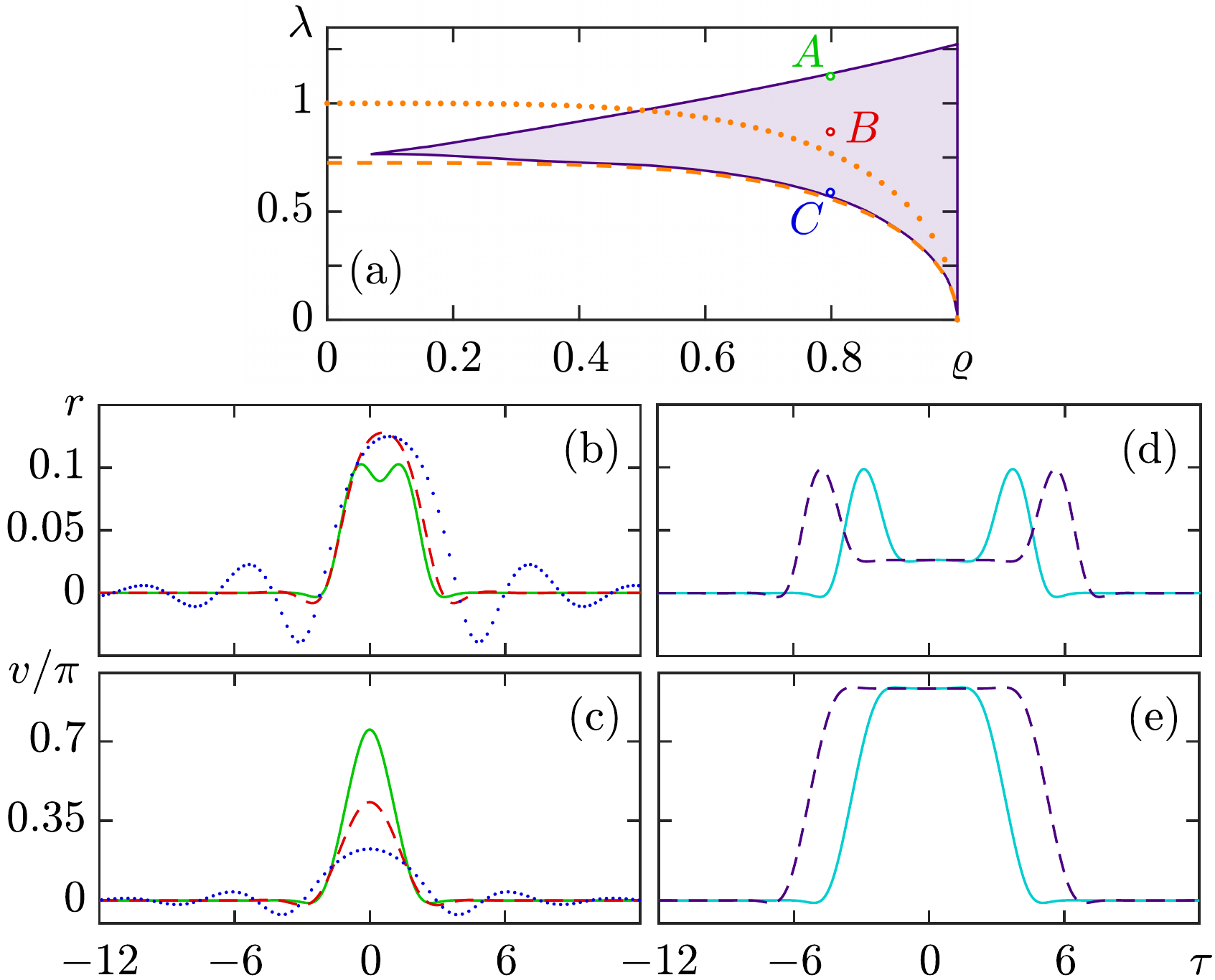}
\caption{(a): Existence region (shadowed) of solitary synchronization waves. 
The dashed and the dotted lines represent the maximal phase and group velocities of the 
linear waves, respectively. (b,c): Solitary waves for 
$\varrho=0.8$ and three different values of the velocity: $\lambda=1.12096$ 
(green solid line - point $A$ in panel (a)), $\lambda=0.85829$ (red dashed line - 
point $B$) and $\lambda=0.58404$ (blue dotted line - point $C$). (d,e): 
Two solitary waves having different widths, moving with practically
the same constant velocity 
$\lambda=1.13952$ on top of a partially synchronous homogeneous background with $\varrho=0.8$.}
\label{fig:slt}
\end{figure}

For general parameters $\varrho$ and $\lambda$ we found solitary 
solutions of system~\eqref{eq:rhotheta} numerically.
With the traveling wave ansatz, the discrete lattice equations~\eqref{eq:rhotheta} reduce
to delay-advanced differential equations for the waveform
$\rho\!\left(\tau\right)=\varrho+r\!\left(\tau\right)$ and $\theta\left(\tau\right)$.
Starting from an approximation obtained analytically as outlined above, we apply an
iterative procedure based on the Newton method to find an exact fixed point of these
equations.
The strategy is to start from solutions very close to synchrony
(i.e. $\varrho\lesssim{1}$), where the shape of the solitary wave is 
known from the perturbation approach, 
and to change parameters
gradually to remain in the convergence domain of the Newton method.
In this way, solitary waves can be found in a large range of parameters, and the
borders of these ranges can be identified, see Fig.~\ref{fig:slt}.
There we also illustrate shapes of solitary waves.
Typical are one-hump $r(\tau)$ profiles for small velocities, and two-hump 
profiles for large $\lambda$ (panels (b,c)).
Tails of the solitons become more wavy close to the lower border, which is 
essentially determined by the resonance with the phase velocity of linear waves.
Close to the top border, the solitons look like  domains bounded by two humps 
(panels (d,e) of Fig.~\ref{fig:slt}).
All such solitons (which are essentially formed by two kinks of 
variable $v\!\left(\tau\right)$ that connect the states 
with $v\!=\!0$ and $v\!=\!v_{\ast}\!\lesssim\!\pi$) share 
nearly the same 
height and the same speed, but their width is not fixed.
This feature is similar to the properties of kovatons  
of the model~\eqref{eq:vnCompactons}.
The maximal group velocity $\lambda^{\max}_{gr}$ of linear
waves is not essential for the existence of solitons, but rather for their visibility.
From a compact initial profile, solitons with velocity larger than
$\lambda^{\max}_{gr}$ dominate the front edge zone;
this occurs for $\varrho\gtrsim 0.5015$.

\begin{figure}[!t]
\centering
\includegraphics[width=\columnwidth]{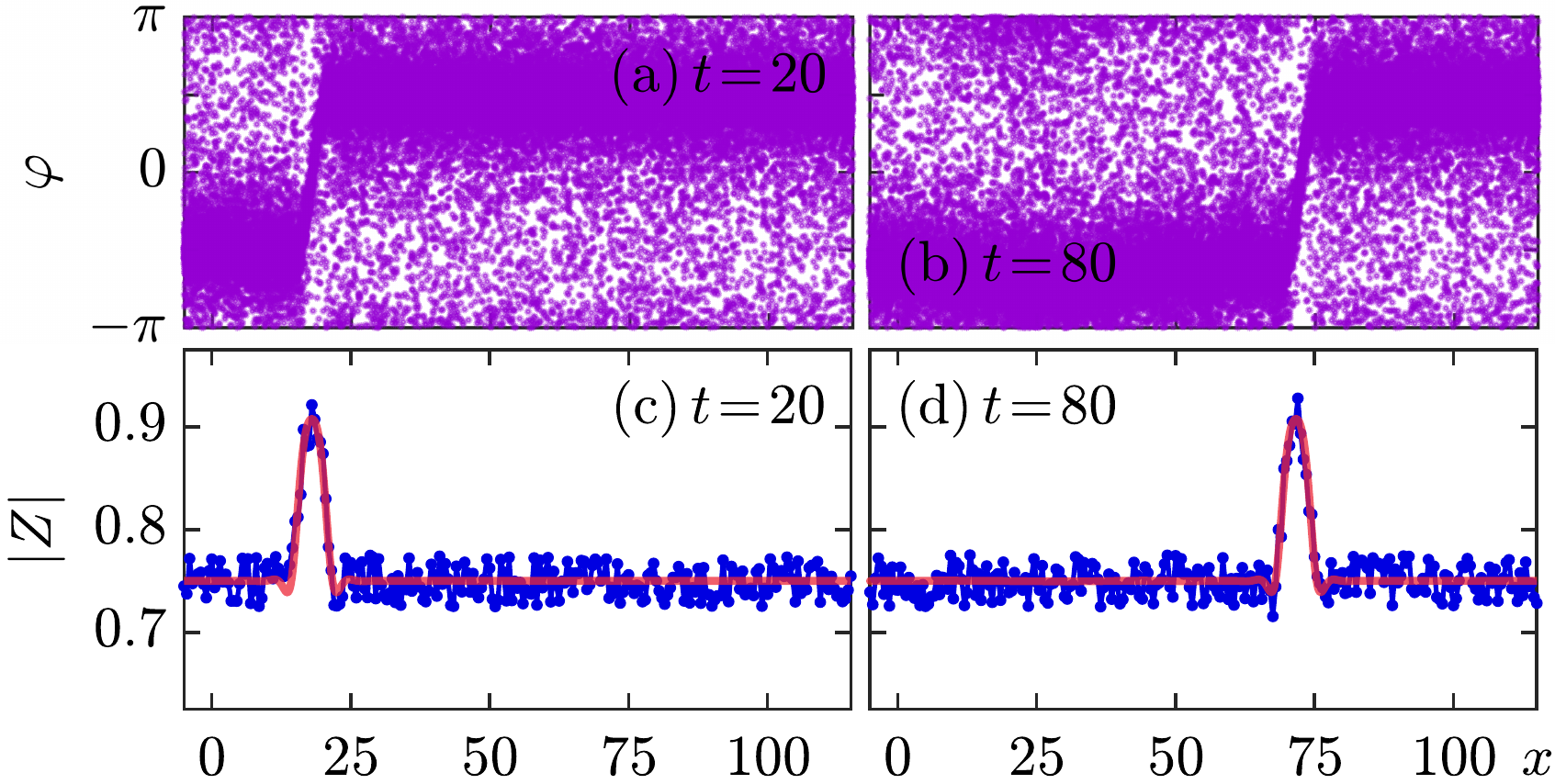}
\caption{Solitary wave simulated in the model~\eqref{eq:phdiscr} with $\ell\!=\!0.00025$ and $G(x)\!=\!2(x^2\!-\!1)e^{-x^2/2}$. (a,b): Instantaneous phases. (c,d): Amplitude of the coarse-grained order parameter, locally averaged with a Gaussian kernel $e^{-q^2/2b^{2}}$ with $b\!=\!100$ (blue line with dots). The red bold line shows the evolution of the corresponding soliton with $\lambda\!=\!0.91$ on top of $\varrho\!=\!0.75$ within Eq.~\eqref{eq:opint}.}
\label{fig:solint}
\end{figure}

Existence of solitary synchronization waves appears to be a general property of
1D media with Laplacian coupling, both discrete ones~\eqref{eq:bm},
and continuous ones~\eqref{eq:opint}. 
To find such waves in Eq.~\eqref{eq:opint}, we have modified
the Newton method and used one of the localized solutions of~\eqref{eq:bm} as a
starting approximation.
We illustrate in Fig.~\ref{fig:solint} the found solitary wave, together with direct
numerical simulations of the original phase model~\eqref{eq:phdiscr}.
One can see that the solitary wave is stable despite the finite-size fluctuations.

\begin{figure}[!t]
\includegraphics[width=\columnwidth]{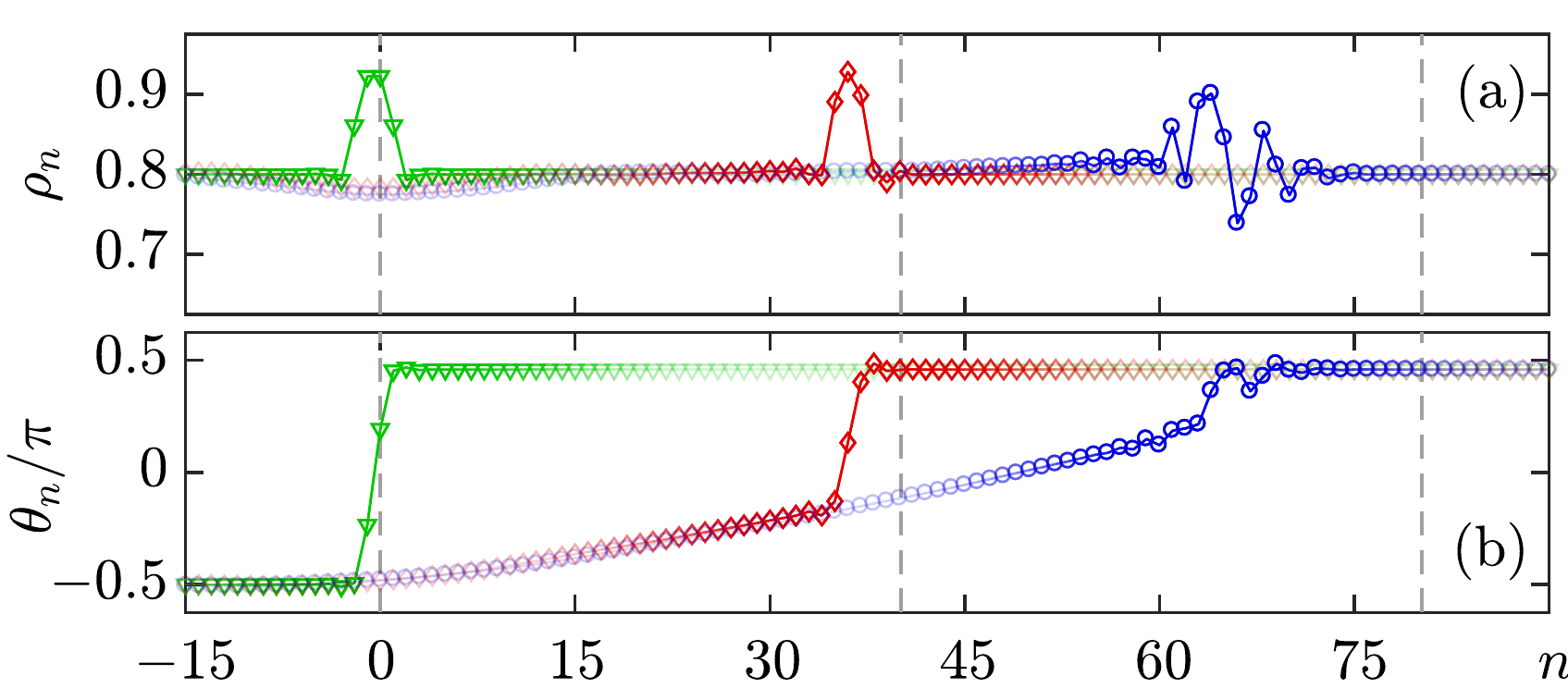}
\caption{Evolution of the initial soliton (green line with triangles), 
calculated for $\varrho\!=\!0.8$ and $\lambda\!=\!0.85307$ in a 
lattice{\,}\eqref{eq:bm}{\,}with slightly non-conservative coupling 
$\alpha\!=\!-0.496\pi$. Profile at $t\!=\!47$ is in red with diamonds, profile 
at $t\!=\!94$ is in blue with circles. The dashed vertical lines show the 
soliton positions if $\alpha\!=\!-0.5\pi$, to illustrate deceleration. 
The amplitude of the soliton decreases and the oscillating tails become visible.}
\label{fig:decsol}
\end{figure}

Above we considered oscillator arrays with purely conservative coupling.
For $|\alpha|\!\lesssim\!\pi/2$ the linear waves decay and one can expect that the
solitons decay as well.
We illustrate this in Fig.~\ref{fig:decsol}.
Here we start with a solitary wave for $\alpha\!=\!-\pi/2$; during the propagation
it gets destroyed. 

Generally, there is another source of synchrony ``non-conservation''.
This is a diversity of oscillators, in particular of their natural frequencies.
It leads to loss of synchrony; in terms of the order parameter evolution, there appears
a dissipative term which reduces the amplitude of $Z\left(x,t\right)$.
To take into account these effects we modify our lattice model 
\eqref{eq:bm} as follows:
\begin{equation}\label{eq:mbm}
\begin{aligned}
\frac{dZ_{n}}{dt}&=-\gamma Z_n+\frac{1}{2}\left(H_{n}-H_{n}^{\ast}Z_{n}^{2}\right),\\
H_{n}&=e^{-i\alpha}\left(Z_{n-1}+Z_{n+1}-2Z_n\right)+\mu Z_n.
\end{aligned}
\end{equation}
Here $\gamma$ corresponds to the assumption that the natural oscillators' frequencies
obey the Cauchy distribution $\pi{g(\omega)}\!=\!\gamma\!\left[\omega^2\!+\!\gamma^2\right]^{-1}$.
The complex parameter $\mu\!=\!\mu_r\!+\!i\mu_i$ describes the level of a local (within the lattice site $n$)
mean-field-type interaction
(in terms of the distributed model~\eqref{eq:phint} this corresponds to an
additional coupling with a narrow mean-field-type kernel).
Finally, the phase shift $\alpha$ can deviate from $\pm \pi/2$, corresponding to an
attractive or a repulsive Laplacian coupling.

\begin{figure}[!t]
\includegraphics[width=\columnwidth]{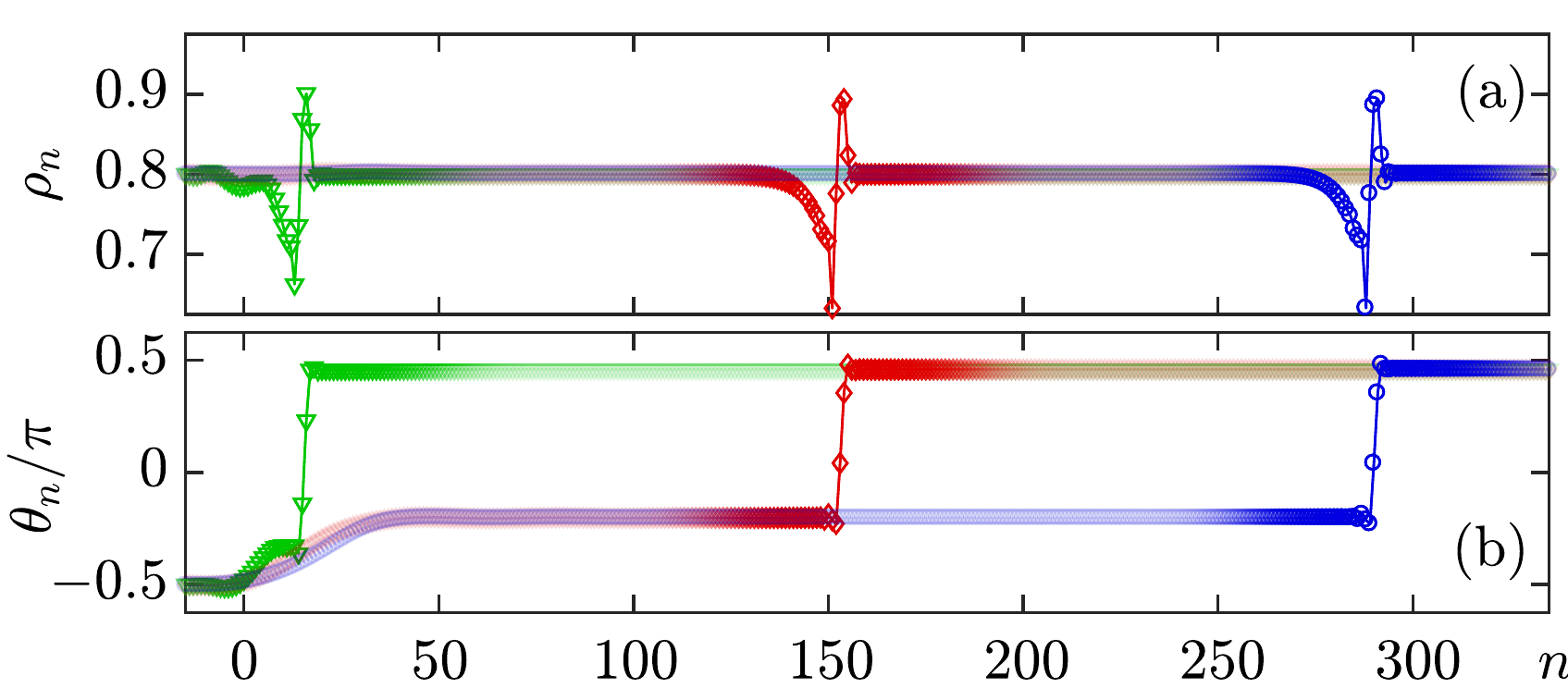}
\caption{Snapshots at $t\!=\!20$ (green line with triangles), $t\!=\!210$ 
(red line with diamonds), and $t\!=\!400$ (blue line with circles) for the numerical 
simulations of Eq.~\eqref{eq:mbm} with $\alpha\!=\!-0.5\pi$, $\gamma\!=\!0.04695$ and 
$\mu\!=\!0.26086\!+\!0.07588i$. Initial conditions: a conservative soliton having 
velocity $\lambda\!=\!0.85307$ and propagating on top of a homogeneous background 
with $\varrho\!=\!0.8$. One can see that a dissipative solitary wave, moving with 
a constant velocity $\lambda_{ds}\!\approx\!{0.72}$, is formed.}
\label{fig:dissol}
\end{figure}

\begin{figure}[!t]
\includegraphics[width=\columnwidth]{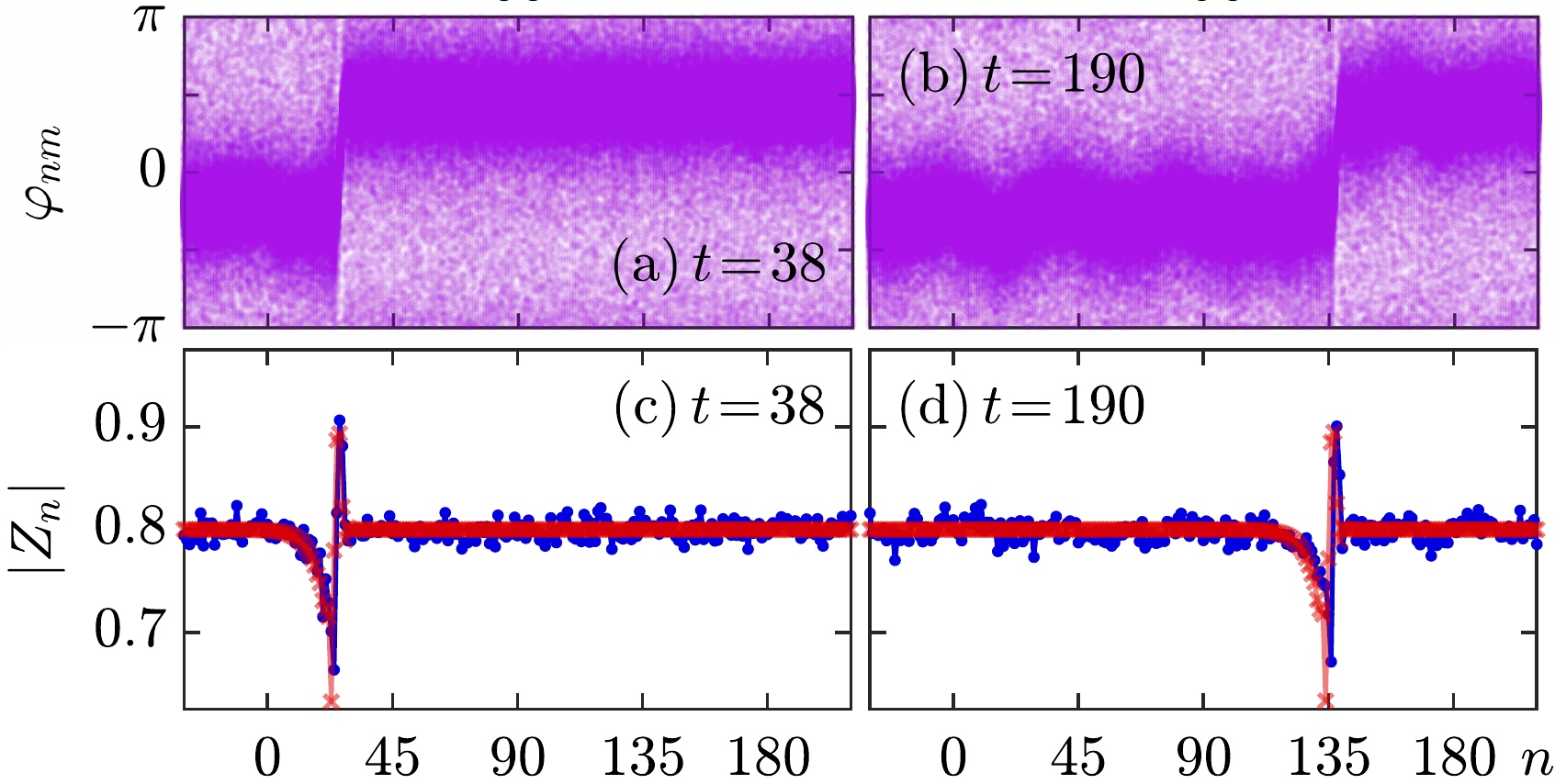}
\caption{Solitary wave simulated in  a chain of $N\!=\!512$ oscillator populations, 
where each population consists of $M\!=\!2000$ coupled elements. (a,b): Instantaneous  
phases. (c,d): Amplitude of the complex order parameter (blue line with dots) compared with solution of~\eqref{eq:mbm} (red line with crosses).}
\label{fig:dis-sol_KSmodel}
\end{figure}

For $\gamma\neq 0$, $\mu\neq0$, only one homogeneous level of synchrony is
possible, given by the stationary solution of Eq.~\eqref{eq:mbm} with
$H_n=\mu Z_n$: $\varrho_{\ast}=\sqrt{(\mu_r-2\gamma)/\mu_r}$.
In Fig.~\ref{fig:dissol} we show what happens to a
localized initial perturbation in such a system.
Here we have chosen parameters $\gamma$ and $\mu$ in such a way that 
the homogeneous state has the same level of synchrony $\varrho_{\ast}=0.8$ as is used in
Fig.~\ref{fig:decsol}, and started with the same initial condition as in Fig.~\ref{fig:decsol}.
After an initial transient, this solution evolves into a localized wave
which is not similar to the conservative soliton, but nevertheless appears to be
stable and propagates with a constant velocity and a permanent form.
This solution can be attributed as a dissipative solitary synchronization wave.
In Fig.~\ref{fig:dis-sol_KSmodel} we also show with direct
numerical simulations, how the found dissipative soliton of the complex order parameter
propagates in a chain of the interacting communities of coupled nonidentical
oscillators where the evolution of the phases $\varphi_{nm}\left(t\right)$ of
oscillator populations at each site $n$ is subject to a mean-field force from the same
lattice site, and to the Laplacian forces from the neighboring sites.
One can see that the solitary wave is stable despite the finite-size fluctuations.


In conclusion, we described solitary synchronization waves in an array of
oscillators with Laplacian coupling.
These waves are propagating with a constant velocity profiles of the complex order
parameter; they can be characterized as kinks of the global phase, within these kinks
the local synchronization level is higher than in the surrounding background.
In the limit of a fully synchronized background, only the phase kinks remain, which
coincide with previously studied phase compactons and kovatons.

\looseness=-1
We have presented solutions for the simplest lattice model, and have demonstrated that
they are also stable in large populations with integral coupling terms.
For identical oscillators with conservative coupling, there is a family of solutions
with different velocities on different backgrounds, similar to other conservative
nonlinear wave systems like the NLS lattice.
For nonidentical oscillators, a finite level of synchrony can be maintained by
attractive coupling; here the solitary waves are dissipative solitons with a certain
amplitude.
A more detailed analysis of the dissipative case will be presented elsewhere.

While we focused just on solitary waves in this Letter, we can mention that general
initial conditions typically lead to rather complex, turbulent patterns of the 
order parameter.
If the initial profile is a localized bump on a constant background, typically at 
the propagating edges a system of solitary waves is formed, and at large times the leading
soliton with the largest velocity is well separated from the waves behind it.
We, however, have not studied interactions and collisions of the solitary waves.

\acknowledgments{
We thank P. Rosenau and A. Nepomnyashchy for fruitful discussions.
L.A.S. thanks DAAD (Grant N. 91697213). 
The work was supported by Russian Scence Foundation (Grant N. 17-12-01534).}

\end{document}